*Article*

# Where Photons Have Been: Nowhere Without All Components of Their Wavefunctions

**R. E. Kastner**

(University of Maryland, College Park; rkastner@umd.edu)

**Abstract**

A nested interferometer experiment by Danan *et al* (2013) is discussed and some claims evaluated concerning the whereabouts of the photon, primarily in the context of time-symmetric interpretations of quantum theory including the Two-State Vector Formalism (TSVF) and the Transactional Interpretation (TI). It is pointed out that the TSVF account fails to predict the observed data based only on the first-order wavefunction component. It is shown that the Transactional Interpretation readily accounts for all the observed phenomena.

## 1. Introduction

This paper evaluates suggested implications of an experiment presented by Danan *et al* in a paper entitled "Asking Photons Where They Have Been" (Danan, Farfurnik, Bar-Ad and Vaidman, 2013; henceforth "DFBV") [1] and offers another way of understanding the experimental results. There has been much discussion of this paper: e.g., [2] Li, Z. *et al* (2013), [3] Svennson (2014), [4] Salih (2015), [5] Danan et al (2015), [6] Griffiths (2016), [7] Sokolovski (2016), [8] Nikolaev (2017), [9] Englert, B.-G. et al. (2017), [10-12] Vaidman (2013, 2017a,b), [13] Wiesniak (2018)). I will argue that the interpretation of the observed data by way of the "Two State Vector Formalism" (TSVF), cf. [14] relies too heavily on a truncated wavefunction, and that the entire wavefunction of the prepared photon beam is required for a correct account of the experimental results. In an Appendix, I analyze the experiment in terms of the Transactional Interpretation (TI), which takes into account the entire quantum state of the photon beam.[1]

The basic experimental setup presented in DFBV consists of an outer interferometer within which is nested an inner interferometer. The experiment's "surprising" data are obtained by tuning the inner interferometer so that there would be destructive interference at its exit point, were it not for small vibrations of the mirrors in the inner interferometer that give rise to some leakage and also to the observed signals.

Mirrors at various points within the interferometers are set to vibrate with distinct frequencies as a means of correlating the transverse degree of freedom of the photon beam with the longitudinal degree of freedom traversing the various arms. Signals arising from the scatter of the transverse degree of freedom are then obtained. Below are replicas of figures presented in [1] (henceforth DFBV), showing different configurations of the experimental setup along with the signal data correponding to each configuration. However, it should be noted that the figures omit crucial mirror leakage. This makes them (and the

[1] TI has acquired some additional nomenclature based on its recent development into the relativistic domain. For consistency in the present work, in what follows I will generally use the acronym "PTI" for "Possibilist Transactional Interpretation," since that is the usage in a passage fron Stuckey quoted in the text.

arguments making use of them) at best tendentious. (This issue is addressed in some detail in what follows.)

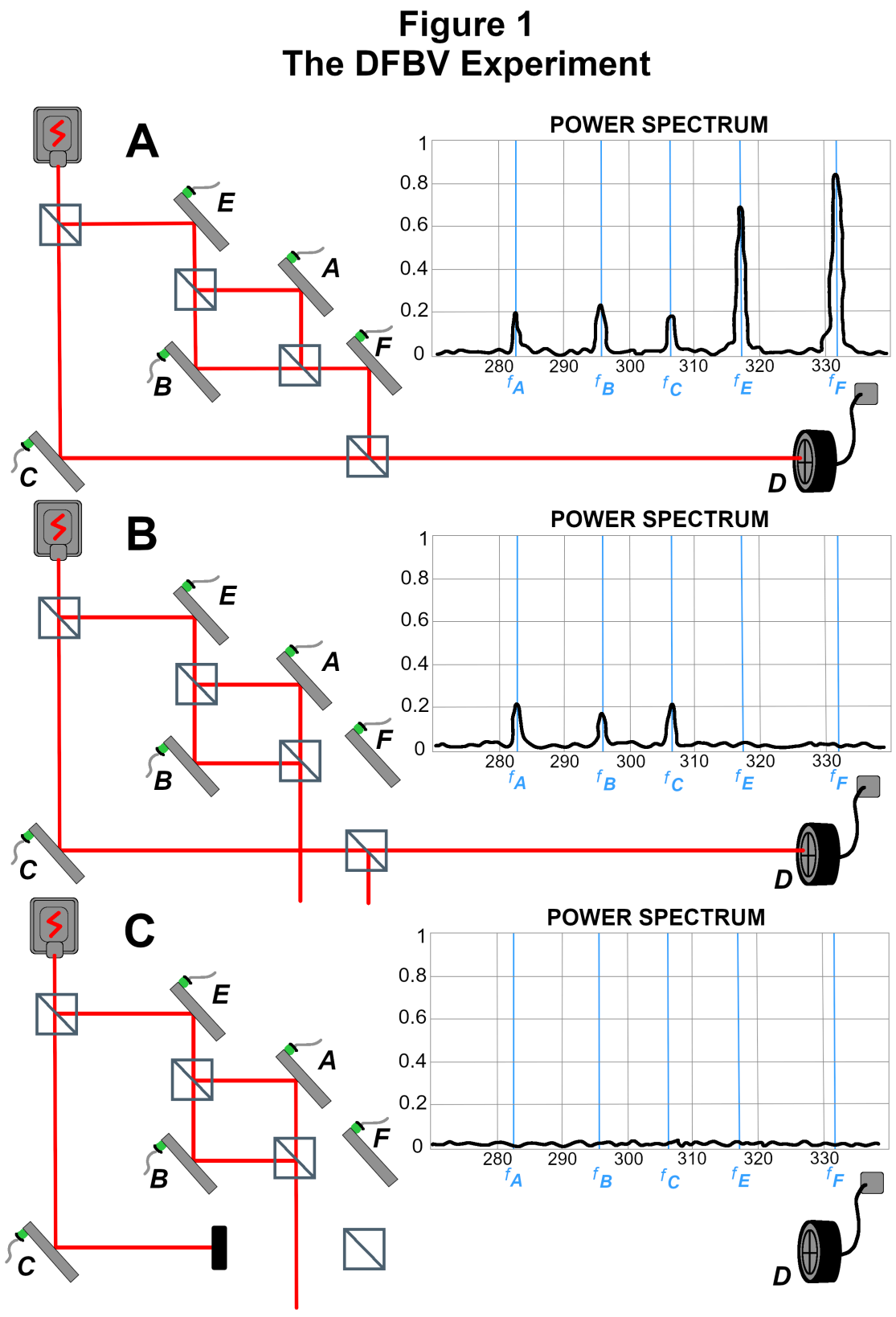

**Figure 1.** The DFBV experiment showing only longitudinal degree of freedom (redrawn from DFBV 2013).[2].

In Figure 1, the inner interferometer can be tuned for constructive or destructive interference at the exit point of the inner interferometer. Mirrors A, B, C, E, and F each vibrate at their own frequency with a small amplitude so as to constitute a form of 'weak measurement' of the beam component encountering each mirror. The signal comprising the weak measurement is obtained by picking up the specific frequency of each mirror's vibration as revealed in the scatter of detections, where that scatter arises from variations (created by each mirror's vibration) in the transverse wave function component. The inner interferometer can be configured for destructive interference so that the longitudinal compoment that exits the inner interferometer toward mirror F almost cancels out, except for leakage due to the mirror vibrations. This case is illustrated in Figure I B (but recall that the red depicted "path" of the beam omits the longitudinal leakage, as well as the transverse wavefunction component, both which vanish nowhere in the apparatus).

DFBV apply the Two-State Vector Formalism (TSVF), together with 'weak values' of observables, to the experiment as their proposed way of accounting for the phenomena, since they also used TSVF as a heuristic aid to conceiving the experiment. In TSVF, a quantum system is described by a constructed formal object called a 'Two State Vector,' or TSV, which consists of the preparation ket $|\Psi\rangle$ to the right of the measurement outcome brac $\langle\Phi|$, i.e.: $\langle\Phi||\Psi\rangle$. The 'weak value' of observable O is defined as $\langle O\rangle_W = \frac{\langle\Phi|O|\Psi\rangle}{\langle\Phi|\Psi\rangle}$. Note that the TSV "$\langle\Phi||\Psi\rangle$" has two interior lines and is not an inner product. The founders of the TSVF emphasize that its predictions deductively follow from

2 All figures rendered by Wendy Hagelgans.

standard quantum theory and present TSVF as "a time-symmetric description of the standard quantum mechanics" ([15] Vaidman, 2007).

The authors acknowledge in their discussion that the signal in 1B is made possible by leakage (omitted from their figures) from the inner interferometer.[3] They also note that the results can be accounted for in standard one-vector quantum theory by the same analysis as that for classical electromagnetism (DFBV, Appendix) but argue that TSVF is a more "elegant" way of explaining the phenomena.[4] We will not repeat that discussion here, but refer the reader to the original DFBV paper or to Stuckey's instructive account [16, 17] (2015a,b).

## 2. Standard Quantum Mechanics Account

Saldanha (2014) [18] elaborates the standard one-vector account, arguing for the efficacy of that formulation. We briefly review that analysis here. The transverse y-component of the beam is a Gaussian envelope; only that y-component is affected by the mirror tilts. In momentum space it is given by:

$$\Psi(k_y) = Ne^{-\frac{k_y^2}{\sigma^2}} \tag{1}$$

Given the above, the transverse components after the beam splitter and following mirrors *{i}* are of the form:

$$\Psi_i(k_y) = \frac{1}{\sqrt{3}}\Psi(k_y - \kappa_i) \tag{2}$$

where $\kappa_i$ are the deflections in the transverse momentum $k_y$ due to the instantaneous tilt of mirror *i* at time *t*.

The transverse beam component reaching the final detector D at any particular time *t* is a superposition of the transverse component from mirror C and from mirror F, namely:

$$\Psi_D(k_y) = \frac{1}{\sqrt{3}}\Psi_C(k_y) + \sqrt{\frac{2}{3}}\Psi_F(k_y) =$$
$$\frac{1}{3}[\Psi(k_y - \kappa_C) + \Psi(k_y - \kappa_E - \kappa_A - \kappa_F) - \Psi(k_y - \kappa_E - \kappa_B - \kappa_F)] \tag{3}$$

Saldanha shows that there are many values of the $\kappa_i$ , *i* = *A*, *B* for which a clear transverse signal is obtained from the inner interferometer, but that no signal is obtained for *i* = E, F. Intuitively, the latter comes about because in the case of destructive interference, the contributions from mirrors E and F always cancel each other out, while those of A and B do not. This can be seen explicitly by using the fact that each $\kappa_i \ll \sigma$ and evaluating (3) to first order. One obtains[5]

---

[3] Thus, the assertion in Vaidman (2017b) that "The experiment shows a disconnected trace left by the the photons in the interferometer" is not supported by the physics, in which there is no actual disconnection at the level of the wavefunction (as acknowledged by DFBV in their original paper).

[4] DFBV make use of weak values as part of their ontological account of the experiment, but care is in order in this regard. Ontological inferences regarding 'strange' weak values and/or counterfactual usage of the ABL Rule [19] (Aharonov, Bergmann and Lebowitz 1964) are a matter of some controversy in the literature (e.g., [20] Bub and Brown (1986), [21] Mermin (1997), [22-25] Kastner 1998, 2003, 2004, 2010 and references therein). Stuckey [17] argues that weak values with second-order contributions lack the kinds of ontological implications often attributed to them.

[5] Here we use (1) and (2) for the various components of $\Psi_i(k_y)$, $e^x \sim 1 + x$ for x<< 1, and drop squared terms in $\frac{\kappa_i}{\sigma}$ .

$$\Psi_D(k_y) \approx \frac{1}{3}\Psi(k_y - (\kappa_C + \kappa_A - \kappa_B) \quad (4)$$

Thus, the dependencies on the tilts from mirrors E and F cancel out and we are left only with those of A,B, and C, which contribute in equal amounts albeit with opposite signs for $\kappa_A$ and $\kappa_B$ (see Saldanha eqs. (6) and (7)).

We are reminded by (3) and (4) that *it is the transverse degree of freedom alone* that contains the signal information from the mirrors, not the longitudinal one, which acts only as a carrier of the information-bearing transverse degree of freedom. Thus, for values of $\kappa_A$ and $\kappa_B$ foiling full destructive interference at mirror F, the contributions of mirrors A, B, and C are essentially equal, and the small longitudinal component from mirror F is sufficient to convey the transverse component to the final (unlabeled) beam splitter where it recombines with the substantial longitudinal wave from mirror C and proceeds to detector D. A small longitudinal component from F does not equate to a small or missing signal as long as enough is present to convey the transverse component to the final beam splitter to recombine with the contribution from C, which is what we get in (4) for $\kappa_A \neq \kappa_B$.

Stuckey's Figure 5 [16], reproduced here as Figure 2, presents an adapted version of Figure 1B showing the transverse component riding on a very small but sufficient longitudinal component to connect the source to detector D with the help of the contribution from C:

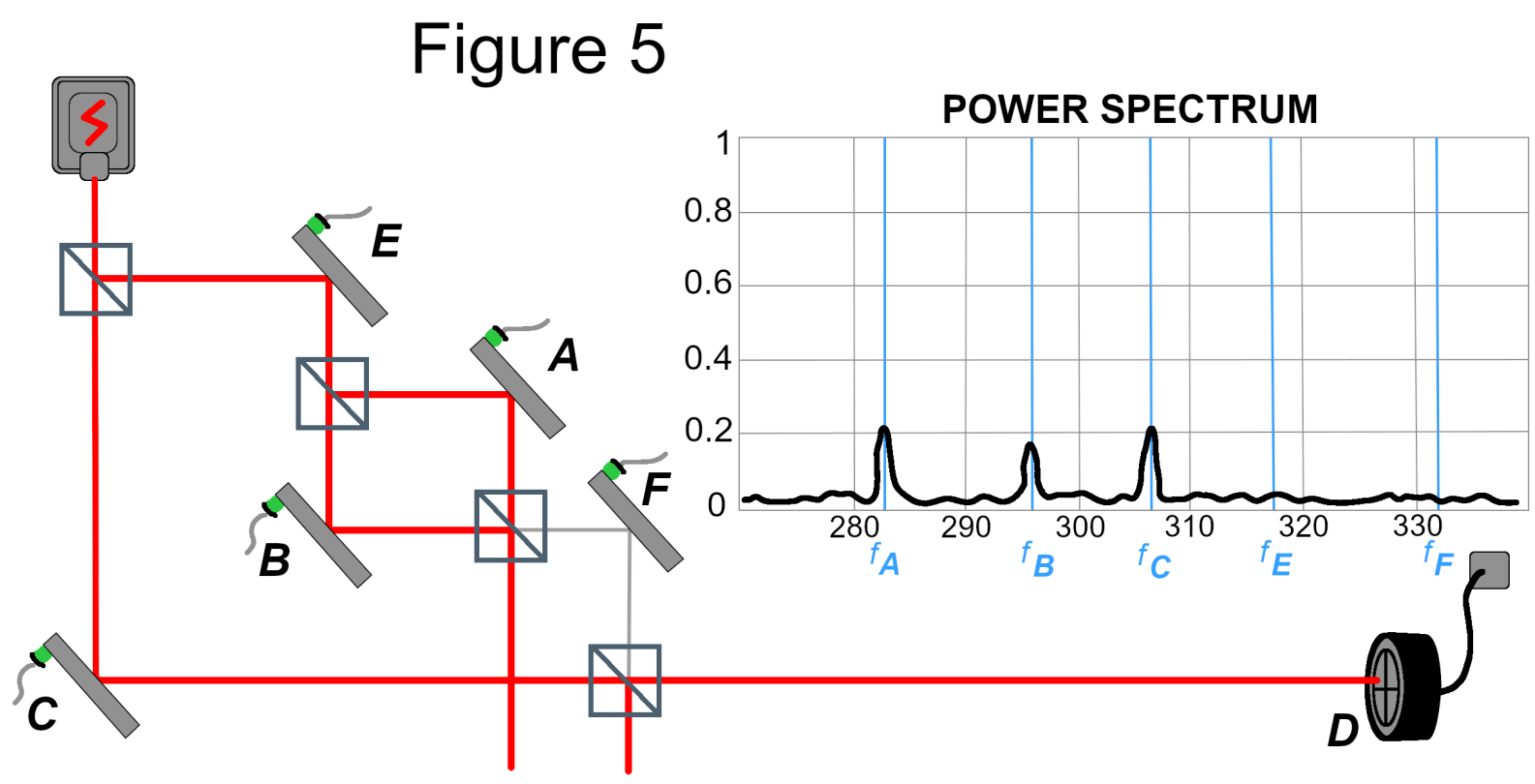

**Figure 2.** The case of destructive interference showing the tranverse component (in gray) being conveyed to the final beam splitter.

This is where the signal comes from, even though it may seem counterintuitive that we get such a strong signal despite such a weak longitudinal component from mirror F. However, we should not let our intuitions get in the way of the unambiguous theoretical expressions (e.g., eqn. 4) that instruct us that the signal-bearing transverse wave function amplitude is not a function of the longitudinal wave function amplitude (for values of $\kappa_i$ yielding signals). As Stuckey points out, the strength of the signal results from the ability of the contribution from mirror F (however small the amplitude of the longitudinal component) to superpose with the main beam from mirror C such that its transverse component is conveyed to detector D.

Thus, the above analysis provides an explanation for the counterintuitive features of this experiment from within standard one-vector quantum theory. The essential point is that one does not actually need to "build up" the longitudinal component at all, since even a small leakage from the "blocked" mirror is enough to recombine with the robust longitudinal component from C to convey the transverse component's signal to the final detector.[6]

## 3. Critique

Danan *et al* argue that the heuristic utility of considering the post-selection in devising of experiments such as the above is evidence for the ontological correctness of the TSVF description. They also argue, based on their depiction of the photon's wavefunction as involving a gap between F and D, that a backward-propagating state is needed to 'build up' the longitudinal component corresponding to mirror F for the case of destructive interference. This idea is represented by the green dashed line in DFBV's Figure 1E (reproduced here as Figure 3). However, as noted above, this argument is available only by neglecting the transverse component and the non-vanishing leakage from F. In particular, the analysis above shows that there is actually no need to "build" anything up beyond the standard quantum state, since it is indeed continuous past mirror F, even though small (see footnote 6), and the longitudinal component from C is sufficient to convey the transverse component's signal to the detector. Moreover, and importantly, were there actually a physical gap in the wavefunction there would be no signals from mirrors A and B. Thus the modeling of the photon's state as a truncated longitudinal wavefunction (with or without a backward-evolving post-selection wavefunction) is neither necessary nor sufficient to account for the data obtained.

This last point is crucial and is worth examining in more detail. Leaving aside interpretational issues of what is to be regarded as the "path of the photon," *the TSVF quantitative analysis itself fails to predict the observed signals from mirrors A and B based only on the first-order longitudinal component.* This is because those signals are generated by the transverse component, and the only way the transverse component terms for A and B get to D--to be post-selected--is by way of the second order longitudinal leakage. In more specific terms, recall that the detected data is a Gaussian envelope in the transverse y-direction with a distinct scatter induced by each mirror's vibration. Thus we cannot disregard the fact that "detector D" comprises a large group of pixels, and each such pixel is a post-selection site for TSVF. To properly apply TSVF to this situation, one cannot disregard this finer configuration but must allow for post-selection at all the pixels comprising the Gaussian envelope. Taking that into account, but using only the first-order wavefunction component as in DFBV, TSVF itself predicts that photons would not be detected at pixel groups corresponding to the scatter induced by mirrors A and B. Thus no photons would be post-selected in those states either. Thus, there is no information from A and B in either the pre- or post-selected states, so the TSVF model using the truncated wavefunction fails to predict the observed data. Instead, it predicts a signal from mirror C and no signals from mirrors A and B. This alone should disqualify the authors' TSVF account using only the truncated wavefunction.

Now of course DBFV claim that the prediction can be based solely on weak values for the truncated (first-order) longitudinal wavefunction, all of which are 1/3. However, a physical modeling and analysis of an experiment must predict the behavior of the pointer, which in this case is also an aspect of the photon itself. As shown above, without the leakage, the photon-pointer would not yield signals from A and B consistent with those weak values, and the experimental results would falsify, rather than confirm them, as a property (or putative "trace") of the photon. This situation illustrates that we can't really "have it both ways", i.e., we can't use information obtained from a weak measurement without paying the price for that measurement, which in this case is the leakage, and that price needs to be included when considering ontological implications of weak measurements. Specifically, one does not get to claim that photons were present in "discontinuous paths" when actually modeling a photon by the discontinuity (truncated wavefunction) fails to predict the signal that is cited as evidence of their presence in discontinuous paths.

[6] There is a Native American legend describing how a tiny sparrow returned once-lost celestial music to the forest by hiding in the neck feathers of a huge eagle who flew up to try to retrieve it. The eagle was unable to make it all the way up to the celestial heights necessary, and fell back, but at that point the sparrow jumped out and flew the rest of the way, succeeding in its mission. This little parable seems to describe what happens in this experiment: the "sparrow" (leakage) is crucial in that its presence or absence dictates the success or failure of the mission. The fact that the sparrow is small does not negate its importance. It only has to be nonzero. A deeper insight, perhaps, is that the difference between zero and non-zero is a "digital" one; when it comes to wavefunctions, a "small" quantity cannot necessarily be considered negligible and disregarded by reference to first-order approximations.

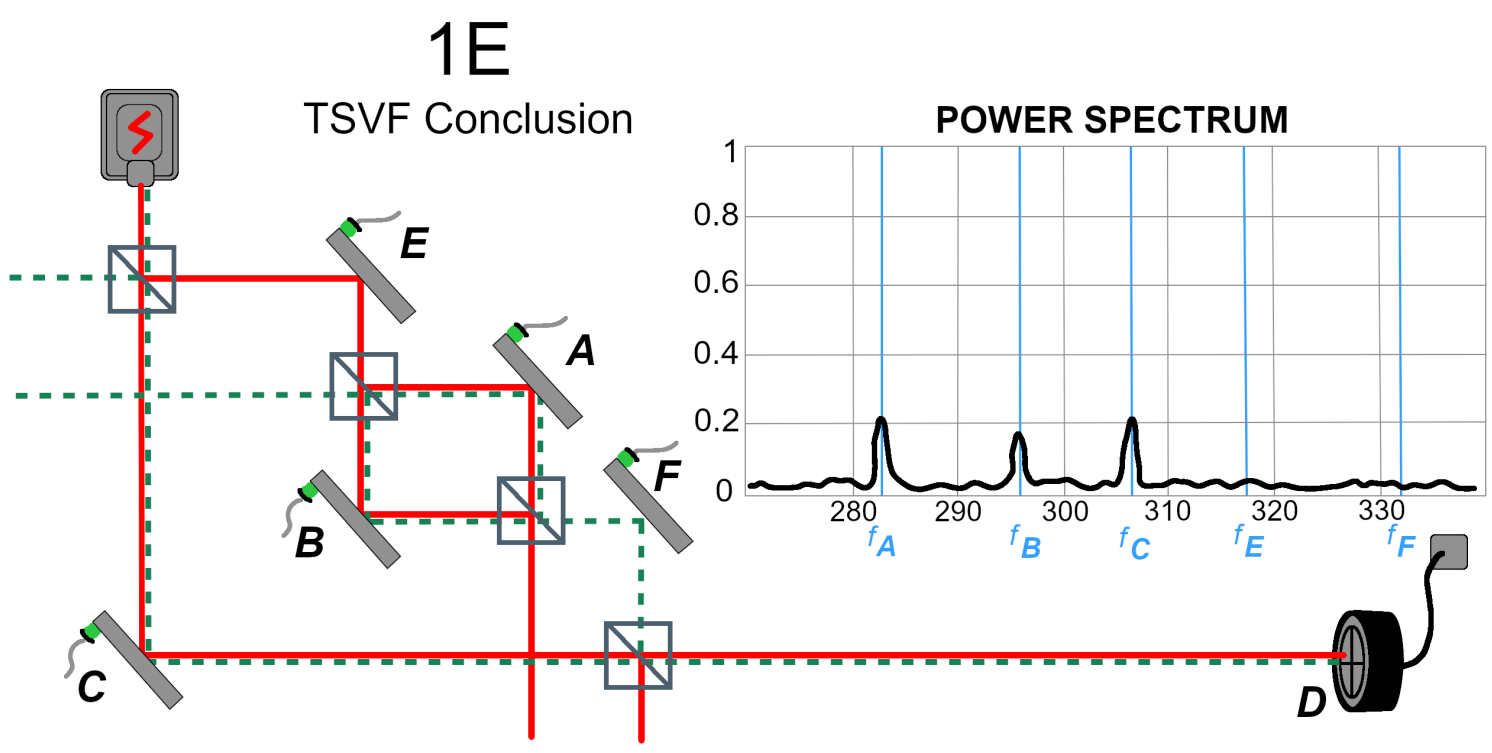

**Figure 3.** The proposed retrocausal post-selection influence in DFBV's experiment. Signals are shown from A and B that would not be present given the depicted discontinuities in the red and green wavefunctions.

In this regard, it is concerning that the authors' Figure 1E (and 1B) show signals from A and B that would not actually be there if the depicted wave function discontinuity existed. (See Figure 3.) The signals are outcomes of the pointer observables correlated with arms A, B, and C, which we could call the "Mirror" observables: $M_A$, $M_B$, $M_C$ respectively applying to the transverse wave function (pointer). The truncated wavefunction component depicted fails to reach D, so the only photons post-selected at D are those whose transverse component reflects a non-zero result for $M_C$.. Consequently, though the backward-propagating green wavefunction has a longitudinal component for mirrors A and B, it has no transverse terms corresponding to signals for mirrors A and B. Thus under the depicted wavefunction, TSVF itself predicts no signals from mirrors A and B, contrary to the data shown.

The ability of the small leakage from F to superpose with the strong longitudinal component from arm C such that one obtains full signal visibility from arms A and B may seem counterintuitive, but then the question arises: is it appropriate to address a counterintuitive aspect of quantum theory by truncating the full wavefunction so as to suppress the content that explains that counterintuitive aspect (as is done in the Danan *et al* figures above)? Is omitting that content (i.e., the second degree of freedom and the leakage that gets it where it needs to be) really a "more elegant" explanation, or is it reliance on an approximation that neglects crucial physics in order to retain a conceptual picture that is undermined when the omitted content is included? (Not to mention the fact that the omitted content is *required* in order to actually predict the observed data, as discussed above.) These questions are raised not out of any *a priori* opposition to the TSVF approach, but because they go to the heart of the relationship between experimentation, physical theory, and ontological conceptualization. If experimentation is to reliably guide the latter, arguably one must guard against presuming a particular ontology to be correct and tailoring the theoretical description of an experiment to provide an optimized fit for that preferred ontology. While the heuristic utility of considering post-selection is evident, the concern here is that we need to be wary of equating heuristic utility with ontology, especially when one retains such a correspondence only by disregarding crucial wavefunction components. There are other reasons to consider post-selection as ontologically important that do not involve omitting crucial parts of the quantum state. These include experiments such as the delayed-choice experiment (DCE).

Before turning to that topic, let us summarize the main concern: In this experiment, manipulating a second degree of freedom (the transverse component of the photon wavefunction) to couple the photon with mirrors A, B and C is what allows the experimenters to create a tomography of the prepared state, whose amplitudes associated with arms A, B and C are revealed in the transverse displacements of the beam encoded in $\Psi(y)$. That is why omitting the second degree of freedom, as well as the leakage from F in the figures and in much of the discussion, is inappropriate. In effect, the figures omitting the crucial leakage create the illusion of a need for the backward-propagating wave as an ingredient of the signal amplitude. But the physical analysis shows that there is, in fact, no such need. The very manipulation employed to obtain the information is what "unblocks" the ostensibly blocked path from A and B to D. And again, as noted above, there one must explicitly include the leakage, since without it, no signals are predicted from A and B.

Perhaps a more fundamental consideration is that the alleged "location" of the photon at particular mirrors at particular times implies the modeling of photons as wave packets. However, photons are not really localizable in this manner in principle, due to the uncertainty relation (see, e.g., [26] Bialynicki-Birula, 2009).

An additional remark is in order concerning the tacit assumption in DFBV that the small amplitude of the forward-propagating longitudinal component in the internal interferometer at F is to be equated to a "small number of photons" (or even zero, as implied by their diagram 1B omitting the wavefunction leakage). In effect, this claim implicitly depends on the idea of a counterfactual measurement at mirror F. That is, indeed if there were a detector at F instead of a mirror, very few photons would be detected there. But also if there were, we would get no signals from A and B, yet we do. (As noted above, this cannot be predicted from the truncated wavefunction, so it will not do to say "that is why we need the post-selection state.") This is a reminder that the entire, actually instantiated experimental arrangement must be taken into account; one cannot 'chop up' an experiment into separate pieces and consider them separately to draw ontological conclusions. In fact there is no detector at F, and the non-vanishing forward-propagating wave can indeed engage in constructive interference beyond mirror F with the C-arm component, producing the observed signals.

Thus the notion that one needs to "build up" the forward-propagating wave with a backward-propagating wave (as depicted in Figure 3)—because otherwise there wouldn't be enough photons in the inner interferometer—implicitly depends on a particular strong claim regarding the ontological significance of counterfactual measurements (as well as an assumption that one can consider experimental components separable). Such claims are widely disputed in the literature based on the ill-defined nature of the 'pre- and post-selected ensembles' involved--in short, this is due to the inability to keep fixed the probabilities for the post-selection result (see also note 4). But in any case, again, the truncated wave function would not yield the signals whether or not the backward-propagating state is included. Wiesniak [13] further notes, "introducing the notion of secondary presence (in [12]) does not seem to contribute to explaining these effects within TSVF, and electrodynamics does not give any foundations of such a gradation." In addition, the uncertainty principle contradicts the idea that photons are ever really localized at particular points at particular times, as noted above.

## 4. The Relevance of Post-Selection

Nevertheless, is there some validity to the idea that post-selection influences what happens to a quantum system in the process of measurement? First, we should remark that apart from explicitly time-symmetric approaches, Wheeler's Delayed Choice Experiment (DCE) already suggests that the behaviors of quantum systems at some time $t$ depend on how they are measured at some later time $t + \Delta t$. Specifically, a photon emitted into a which-way apparatus will yield the appropriate distribution of outcomes for a both-ways or which-way measurement regardless of whether the choice is set before or after the time that the photon presumably enters the two arms of the apparatus. Standard one-vector quantum theory provides no specific ontological explanation for this finding other than Wheeler's famous comment that "no phenomenon is a real phenomenon until it is an observed (or registered) phenomenon" [27] (Wheeler, J. A. , 2004). This then begs the question of what counts as an "observation," which cannot be accounted for from within the conventional theory.

The Transactional Interpretation, [28] (Cramer 1986), [29] (Kastner 2022) answers this question and provides an ontological account of situations such as the DCE. It is based on the direct-action (absorber) theory of fields, in which the transferred photon is essentially a combination of the forward-propagating wave function ("offer wave" or OW) and the advanced confirmation from the detector/absorber ("confirmation wave," CW); thus, both are required for actualization and detection of a photon. In this regard, TI has some common ontological ground with TSVF, which also assumes that measurement (or 'post-selection' when specific measurement outcomes are considered) involves a retrocausal influence due to the post-selection state. Stuckey discusses the commonality between basic TSVF and Cramer's original version of TI, both of which seem to require a block world, and then goes on to explore dynamical variations of these approaches:

This leads us to the third type of retrocausal explanation characterized by Elitzur & Dolev's belief[14] "that genuine change, not static geometry, is reality's most basic property," so that "Perhaps, rather, spacetime itself is subject to evolution." Kastner[15] embraces a similar notion with her growing causet version of TI called the Possibilist Transactional Interpretation (PTI). The reason these approaches are retrocausal is that the evolution of the blockworld or the growth of the causet doesn't proceed strictly from past to future, but the present along any worldline evolves, in part, according to future boundary conditions that allow for intervention. In PTI, the offer and confirmation waves exist in "pre-spacetime" and a spacetime network/graph grows out of actualized transactions in a robust sense. Kastner writes[Kastner (201)]: "The possibilist transactional picture can be viewed as a physical basis for the emergence of the partially ordered set of events in the causal set formalism. This formalism is currently being explored as a means to constructing a satisfactory theory of quantum gravity, and it has much promise in that regard. However, even apart from general relativistic considerations, the formalism breaks new ground in showing that, contrary to a well-entrenched belief, a block world ontology is not required for consistency with relativity. The causal set structure is a 'growing universe' ontology which nevertheless preserves the relativistic prohibition on a preferred frame."

PTI and TI have the following in common with the TSVF approach: The post-selection is a crucial part of the ontology. The photon is necessarily described by both the offer wave and the confirmation wave (the latter resulting from the post-selection). According to a PTI/TI analysis of the DFBV experiment, the offer wave reaching detector D has two degrees of freedom, as discussed by Saldanha[6], and per that full physical description the forward time-evolving offer wave is connected to both the Source and detector D through the additional transverse degree of freedom (Figure 5). Thus, unlike TSVF and Relational Blockworld, there is no discontinuity of the photon path in the PTI/TI account of the DFBV experiment. (Stuckey, M. [16])

Regarding the alleged discontinuity mentioned above by Stuckey, it should again be emphasized that the full wave function description of the DFBV experiment involves two degrees of freedom, and there is no real discontinuity in the wave function, however small the longitudinal leakage from mirror F; so the TSVF description asserting a discontinuity (as reflected in Figure 1B and in Figure 3 in both temporal directions) is inaccurate––not just in a 'nitpicking' way but in a way that is consequential for prediction of the experimental data, as discussed above.

Returning now to the DCE, it is well known that the preparation of a photon is not the whole story about the behavior of that photon, since in the DCE the photon is always prepared in a 'both ways' state at an initial time $t_0$ but will be detected in accordance with the mode of measurement at a final time $t_F$ , whether that is a 'both ways' or 'one way' measurement. Let us represent the prepared DCE photon state or OW as

$$|\psi\rangle = \frac{1}{\sqrt{2}}\left(|A\rangle + |B\rangle\right) \quad (5)$$

where A and B are the possible ways to get to the final screen (if there is one). This can be understood in TI as follows: the OW is the preparation state $|\psi\rangle$, but in order for photon detection, one also needs a CW, which is the adjoint of the OW component received by any specific absorber. For the case of a 'which way' detection, the final screen is replaced by an absorber assembly (such a pair of focused telescopes) that can only receive the $|A\rangle$ and $|B\rangle$ components separately. Specifically, for that case, the OW components reaching absorbers A and B are:

$$OW_A = \langle A|\psi\rangle|A\rangle = \frac{1}{\sqrt{2}}|A\rangle$$

$$OW_B = \langle B|\psi\rangle|B\rangle = \frac{1}{\sqrt{2}}|B\rangle \quad (6a,b)$$

Thus, the adjoint CW responses for A and B respectively are:

$$CW_A = \langle\psi|A\rangle\langle A| = \frac{1}{\sqrt{2}}\langle A|$$

$$CW_B = \langle\psi|B\rangle\langle B| = \frac{1}{\sqrt{2}}\langle B| \quad (7a,b)$$

These CWs are 'which way' CWs which interact separately with each of the photon OW components in a non-unitary process that is described by the outer product of the matching OW and CW components ([30] Kastner and Cramer 2018, pp. 211-212). The CWs are roughly analogous to the post-selection states $|\Phi\rangle$ in a 'Two-State Vector' $\langle\Phi||\Psi\rangle$, except that TSVs for this case would be just $\langle A||\Psi\rangle$ and $\langle B||\Psi\rangle$, without the amplitude factors. Thus, in TI there is further dynamics as given by the direct-action ('absorber') theory of fields. Specifically, the OW and CW individually undergo attenuation; in the case of the OW, this is as given in (6a,b). The CWs (7a,b) are further attenuated by their overlap with the (advanced) source state, i.e., by a factor of $\langle A|\Psi\rangle$ and $\langle B|\Psi\rangle$ respectively, such that the final amplitude of the OW/CW circuit back at the source expresses the Born Rule. The explicit formal description of this process is given by the outer product of the respective OW and CW components in (6a,b) and (7a,b), which yields a set of weighted projection operators representing *incipient transactions* (IT) for A and B respectively:[7]

$$IT_A = \left|\langle A|\psi\rangle\right|^2 |A\rangle\langle A|$$
$$IT_B = \left|\langle B|\psi\rangle\right|^2 |B\rangle\langle B| \qquad \text{(8a,b)}$$

The weights satisfy the Kolmogorov probability rules and thus are straightforwardly interpreted as probabilities. Thus, the (non-unitary) transition resulting from absorber response can be identified as the von Neumann measurement transition and the weights as the Born Rule (as discussed in Kastner 2022, Chapter 3).

If the chosen final measurement is instead that of a 'both ways' observable, then the detector configuration is some form of final screen whose individual absorbers at positions X receive contributions from both A and B and thus respond with CW of a form matching the both-ways OW, i.e., CWs which access both ways on their way back to the emitter (modulated by the position X of each absorbing screen component). Specifically, the OW component received by each screen absorber X would be:

$$\langle X|\psi\rangle|X\rangle = \frac{1}{\sqrt{2}}\left(\langle X|A\rangle + \langle X|B\rangle\right)|X\rangle \qquad (9)$$

and its CW response would be:

$$\langle\psi|X\rangle\langle X| = \frac{1}{\sqrt{2}}\left(\langle A|X\rangle + \langle B|X\rangle\right)\langle X| \qquad (10)$$

Thus, in this case a set of "both ways" incipient transactions is set up corresponding to outcomes for each value of X, i.e., $|\langle X|\psi\rangle|^2|X\rangle\langle X|$ , with collapse to one 'winning' absorber X'. It is of course an elementary result that we get interference from the factor $|\langle X|\psi\rangle|^2$ based on the sum over the amplitudes for each slit.

In either case, under PTI, we have real physical collapse to a particular outcome, which may be viewed as a form of spontaneous symmetry breaking. That actualization is accompanied by a temporal symmetry breaking as well, such that the real photon constitutes positive electromagnetic energy propagating in a forward-time direction from the emitter to the absorber. This is what allows PTI to lend itself to a dynamical growing-universe picture (as discussed above by Stuckey) in contrast to the block world apparently required by TSVF.[8,9]. In fact, the transactional formulation has now been developed into

[7] These projection operators apply to the photon in the interval between emission and absorption and are relativistically invariant quantities, rather than being elements of an instantaneous direct product space. At the nonrelativistic level, they function as components of the Heisenberg observable *O(t)* applying to the local time of absorption *t*.

[8] I.e., if every system is ontologically described by a TSV, then every system must undergo a "future" measurement yielding a determinate measurement result that propagates backward as the post-selection state. Otherwise, the system can only be described by its preparation state; i.e., it lacks a TSV. In contrast, in PTI, there may be systems (such as an electron in an energy state |E> bound to an atomic nucleus) which never undergo a measurement interaction. In other words, no future measurement result must be presupposed in order to assign a state description to the system.

[9] The present author respectfully differs with Price's view [31] , discussed in Stuckey [16], that a block world allows for a dynamical 'causal story' in either or both temporal directions. If the block world constitutes a static ontology, then a dynamical causal story is *de facto* inconsistent with that ontology. If the intent is that the story is perspectival, based on the assumption that observers are 'moving through' the block, then only one direction of causal flow is possible corresponding to the assumed direction of observer motion. Certainly there is no relative observer motion through the

a full transactional theory of gravity; [32] Schlatter and Kastner, 2023. In the PTI picture, the future is genuinely open and both OW and CW are generated from quantum entities understood as pre-spacetime forms of Heisenbergian potentiae [33]. PTI provides a quantitative account of the transformation of those possibilities into actualities, which is also an account of the measurement transition in quantum mechanics. The actualized photon in this experiment is a connection between its emitter (e.g., a laser) and its absorber (detector component) that has a presence in all interferometer arms. This presence can be straighforwardly read off its OW and CW, as an outer product (as in eqn. 14; see Appendix), without need for postulations about traces that it may have left in any particular location, although of course detecting evidence in the form of those traces is a nice confirmation of this outer product form of the photon that arises naturally in PTI.

## 4. Conclusion.

The interesting experiment by DFBV could be seen as illustrating the heuristic utility of considering post-selection measurement, but its interpretation warrants caution. As noted above, one cannot disregard the nonvanishing leakage from mirror F and the second, transverse degree of freedom, which together provide for the "weak measurement" and without which there is no signal at all. An accurate analysis shows that there is in fact no discontinuity in the photon path under the weak measurement, and that the signal intensity does not depend on the (small) amplitude of the longitudinal component leaking from mirror F but only on its ability to superpose with the component from arm C. Specifically, if the mirror deviations are large enough to provide a transverse signal, then they give rise to a longitudinal leakage large enough to carry the transverse degree of freedom to the final mirror in order to superpose on the ample component from arm C, which acts as a carrier for the signal. Without ample leakage, that is not possible. Thus, figures in Danan *et al* showing a gap in the photon path along with signals from A and B are manifestly incorrect, since there would be no such signals if the illustrated discontinuity between F and D actually obtained. Pointing to the post-selected state as ostensible support for the signals under the assumption of a gap is therefore not sufficient; as noted earlier, the post-selected wavefunction also would have no components from mirrors A and B without the leakage included.

Nevertheless, the heuristic utility of considering post-selection gains a basis also in the transactional formulation. In that context, post-selection corresponds to the generation of an advanced confirmation wave (CW) produced by a particular absorber (detector) for a given offer wave (OW, corresponding to the usual quantum state). The CW is just as important a contribution to the ontology of the detected quantum as is the OW, since they contribute equally to the actualization of a particular quantum of electromagnetic energy/momentum and angular momentum delivered from a source to a detector. However, the transactional formulation always takes into account the entire offer wave, which, as above, has two degrees of freedom. And in this sense it is a more complete description that does not ask us to overlook a crucial nonvanishing wave function component or second degree of freedom in order to support observed phenomena.

Thus, while PTI has in common with TSVF the idea that the ontology of a photon is dependent not just on its preparation state at time $t_0$ but also its interaction with any detectors at some later time $t_F$, PTI differs from TSVF in that the OW $|\Psi\rangle$ and CW $\langle\Phi|$ are entities described by the direct-action picture of fields, which undergo static spacetime block that would give rise to bidirectional temporal flow. It is therefore unclear, at least to this author, in what sense a bidirectional causal story could be seen as allowable within a static block world ontology.

attenuation and interact in such a way as to yield a weighted projection operator corresponding to the actualized photon (as in (14); for details, see Kastner (2022), Chapter 3). The PTI picture also supports the actualization of outcomes in a quantitatively described process corresponding to the von Neumann measurement transition, cf. Kastner 2022, Chapter 3). Thus, rather than stipulating measurement outcomes for arbitrary future times in order to assign a state description (as TSVF must do in the absence of an account of what constitutes "measurement"), PTI makes use of the physical interactions occurring in a given experiment in the present, and on that basis assigns the retarded state arising from the preparation (OW) and the advanced state(s) arising from the mode(s) of detection. These processes together correspond to von Neumann's non-unitary "Process 1" and the advent of a proper mixed state corresponding to a well-defined measurement basis, which then collapses to a particular actualized outcome via spontaneous symmetry breaking (cf. Kastner 2022, Chapter 4).

Also in contrast to TSVF, PTI does not make use of 'pre- and post-selected ensembles' in which the pre- and post-selection outcome are considered fixed while counterfactual measurements are considered for intervening times. This is because such ensembles are not well-defined in view of the lack of a Kolmogorov probability space for all possible outcomes of all possible observables, and because intervening noncommuting measurements make it impossible to actually fix post-selection measurement results.[10]

It may further be noted that the definition of the weak value, $\langle O \rangle_W = \frac{\langle \Phi|O|\Psi \rangle}{\langle \Phi|\Psi \rangle}$ , reminds us that 'weak values' are essentially off-diagonal operator matrix elements normalized by the pre- and post-selection, and as such are essentially two-time transition amplitudes.[11] Thus, one can point to the fact that operator matrix elements are part of standard quantum theory and that one can conditionalize on the post-selection without adopting the retrocausal ontology of either TSV or PTI. Nevertheless, it is pointed out herein that the Delayed Choice Experiment already suggests some form of retrocausal influence in that a photon somehow 'knows how to behave,' at least in the sense that its outcomes conform to the appropriate probability space, in between its preparation and final detection. The time-symmetric models discussed herein provide ways of understanding the nature of that process from a realist perspective. However, we need to be careful not to omit crucial components of the quantum state when seeking ontological understanding.

Regarding the TSVF, the Two-State Vectors, such as. $\langle A||\Psi\rangle$ and $\langle B||\Psi\rangle$ for the 'which way' case, are formal constructs that would (roughly) correspond in the PTI model to an amalgam of the basic OW from the source and the basic CW for each detector, without their respective attenuations, and as such (according to PTI) would not be viewed as a full description of the physical dynamics leading to a measurement result. This issue, including the ability of PTI to provide an account of the measurement transition and the

[10] For specific cautions about ontological claims based on counterfactual intervening measurements, see Kastner 1998, 2003, 2010; Bub and Brown 1986; Mermin 1997. For example, Bub and Brown (1986) note in their Conclusion: "The somewhat curious analogs of contextuality and nonlocality which arise in the statistics of quantum ensembles which have been preselected and post-selected via an arbitrary intervening measurement have their origin in the fact that such ensembles are not well defined without specification of the intervening measurement."

[11] The normalization corresponds to promoting the sub-ensemble corresponding to a particular post-selection result to the entire sample space, which can lead to inferential fallacies based on thinking of these sub-ensembles as "fixed" when in practice they cannot be, and the entire sample space is actually still in play. This is the origin of fallacies involving counterfactual measurements; see, e.g. Kastner 1998, 2010, and the previous note.

Born Rule in terms of a specific theory of fields, is the primary difference between the two approaches.[12]

As noted by Stuckey, the OW for the DFBV experiment includes the transverse component as well, and while there is still significant attenuation, there is no real discontinuity of the OW between mirror F and detector D nor of the CW between mirror E and the source. In addition, PTI does not model photons as 'wave packets'; rather, the photon is the entire actualized transaction, and as such is not a localized particle or packet with a determinate spacetime trajectory. For example, the photon actualized in a 'both ways' two-slit or interferometer experiment is never localized in either slit or arm. It is equally present in both, and not locatable longitudinally either. This is consistent with the inherent uncertainty regarding photon location in both space and time, as referenced earlier (Bialynicki-Birula , 2009).

Finally, of course the DFBV experiment is intended to realize a form of the "Three Box Paradox," in which successful preparation/detection in the given pre- and post-selection states guarantees that the photon is detected with certainty as having traversed either arm A or arm C depending on which intervening measurement is made.[13] However, in this experiment, according to PTI we have not actually performed an intervening measurement of either A or C, since that specifically requires CW corresponding to either A or C, but none are generated in this experiment. Instead, what we have is an attenuation of the beam to the point where what gets to D is *primarily* (but not 100%) a $|C\rangle$ offer wave whose transverse amplitude contains information about the amounts of $|A\rangle$ and $|B\rangle$ present in the prepared state (as conveyed through the leakage through mirror F).

In conclusion, it has been argued that the analysis in DBFV concerning their nested interferometer experiment makes use of an inappropriate truncation of the photon wavefunction, which fails to account for the observed data. Thus the use of the truncated wavefunction in support of their claim that TSVF best explains the data fails to serve its intended purpose, and associated ontological conclusions about the whereabouts of photons remain unsupported as well (in part due to the intrinsic non-localizability of photons in any case). In contrast, it was shown that either standard quantum mechanics or the transactional formulation (TI/PTI) fully account for the data. While interesting questions remain about the underlying ontology of photons and the relevance of post-selection, PTI offers one way of approaching these questions that does not rely on omitting crucial wavefunction components.

**Acknowledgments:** The author is grateful to Mark Stuckey and two anonymous reviewers for helpful comments.

**Funding statement:** This work was produced with no outside funding to report.

[12] For a more detailed account of the physical circumstances triggering the measurement transition, including the relativistic level of the physical interaction and providing the physical basis of the squaring procedure of the Born Rule in TI, see Kastner [34] (2018) and [30] Kastner and Cramer (2018).

[13] Caution is needed in interpreting the Three-Box situation and related constructions, especially regarding counterfactual intervening measurements. See, e.g., Kastner (1998), (2010).

## Appendix: PTI analysis of the DBFV experiment

We provide here an analysis of the DBFV experiment in terms of the transactional formulation (PTI). Here, we abbreviate "offer wave" as OW and "confirmation wave" as CW.

The full (both degrees of freedom) OW component $|\Psi_{y,z}\rangle$ reaching D is:

$$\left|\Psi_{y,z}\right\rangle \approx \left(\varepsilon\left|A\right\rangle - \varepsilon\left|B\right\rangle + (1-\varepsilon^2)\left|C\right\rangle\right) \otimes \Psi(y)\left|y\right\rangle \tag{11}$$

Here, $\varepsilon$ is a very small leakage amplitude, $\Psi(y)$ is the inverse Fourier transform (i.e., position basis) of the amplitude in (4), and $\Psi(y)|y\rangle$ denotes the OW component reaching a specific absorber in the transverse $y$ direction of D.

The basic longitudinal form (neglecting amplitude factor) of the CW generated at D is:

$$\left\langle D\right| = \frac{1}{\sqrt{3}}\left\langle C\right| - i\sqrt{\frac{2}{3}}\left\langle F\right| \tag{12}$$

There is also the basic transverse CW component corresponding to the detection at the transverse absorber element $y$: $\langle y|$. Thus, the basic form of the entire two-degree of freedom CW is $\langle D|\otimes\langle y|$ .

The specific CW (including amplitude factor) generated in response to $|\Psi_{y,z}\rangle$ (given in (11)) is the adjoint of the component of the OW received at absorber element $y$ of D, which is given by the projection of $|\Psi_{y,z}\rangle$ onto $|D\rangle\otimes|y\rangle$ . The generated CW is thus:

$$\left\langle \Psi_{yz}\right|\left(\left|D\right\rangle \otimes \left|y\right\rangle\right)\left\langle D\right| \otimes \left\langle y\right| \tag{13}$$

Given (11), which shows that the longitudinal OW component reaching D is overwhelmingly $|C\rangle$, this reflects the attenuation depicted (misleadingly) by the gap in the red line for mirror F in Danan *et al*'s Figure 1E, shown here as Figure 3. The further attenuation of the CW as it returns to the source is given by the amplitude of the projection of $\langle D|\otimes\langle y|$ onto $\langle\Psi_{y,z}|$, i.e., $(\langle D|\otimes\langle y|)|\Psi_{y,z}\rangle$, reflecting the attenuation depicted (misleadingly) by the gap in the dashed green line for mirror E. The total attenuation resulting from the entire circuit, $\left|(\langle D|\otimes\langle y|)|\Psi_{y,z}\rangle\right|^2$ , is the Born Rule weight of the resulting incipient transaction, which is represented by the outer product of the OW and CW, i.e.:

$$\begin{aligned} &\left|\left(\left\langle D\right| \otimes \left\langle y\right|\right)\left|\Psi_{y,z}\right\rangle\right|^2 \left|D\right\rangle\left\langle D\right| \otimes \left|y\right\rangle\left\langle y\right| \approx \left|\left\langle D\middle|C\right\rangle^2\right|\left|\Psi(y)\right|^2 \left|D\right\rangle\left\langle D\right| \otimes \left|y\right\rangle\left\langle y\right| \\ &\approx \frac{1}{3}\left|\Psi(y)\right|^2\left|D\right\rangle\left\langle D\right| \otimes \left|y\right\rangle\left\langle y\right| \end{aligned} \tag{14}$$

Thus, the contribution to the Born Rule probability for detection at D for various values of $y$ primarily comes from the $|C\rangle$ contribution and is essentially the square of $\Psi(y)$, which yields the observed signals. While we take the first order approximation of the longitudinal degree of freedom in the final step evaluating the Born Rule, the PTI formulation never omits the leakage, without which $\Psi(y)$ would not contain the information from mirrors A and B. However, it should be noted that this fact applies regardless of interpretation. In other words, the first order approximation on the right hand side of (14) that leads to Ψ(y) corresponding to equation (4) is justified *only by acknowledging that the OW from F to D is continuous*. The PTI account of the experiment conceptually diverges from the proposed TSVF account in that respect. And arguably the PTI account is more correct, since it is admitted in Danan *et al* that the leakage is crucial, despite being suppressed in the discussion and figures provided by the authors as they argue that the post-selection state is needed to "build up" the longitudinal component. Yet, we see from the explicit analysis that this is not the case: for given small values of $\kappa_i$ yielding signals, the amplitude of the transverse component is not significantly affected by the attenuation of the longitudinal compoment at F and the leakage conveys it to the final mirror to combine with $|C\rangle$. In any case, as noted above, the actual existence of a gap in the wavefunction (as depicted in the figures in Danan et al) would result in no signals from mirrors A and B .[14] Nevertheless, in the TI picture the confirmation from the detection at D is a crucial aspect of the measurement, in that it is required to create the photon, and also yields the Born Rule probability.

The relevance for retrocausation in the TI picture is as follows: (14) describes incipient transactions associated with different values of $y$, and these give us information about the longitudinal components of the OW (CW) as each interacts with the apparatus. Since PTI views the OW and CW as descriptions of possibility [33], these interactions take place at the level of possibility (pre-spacetime).[15] It is only the 'winning' actualized transactions at a particular value of $y$ for time $t$ that establishes a spacetime process in which a photon is delivered from the source to a specific absorber element $y$ in detector D, primarily via arm C, but with traces left (through the transverse probability $|\Psi(y)|^2$) quantifying the possibility of being associated with A or B based on the prepared state. As in the DCE, in which the detection 'reaches into the past' to bring about the complete process associated with a given measurement, the absorption event at D is actualized at time $t$ while the emission event is actualized at an earlier time, $t-\Delta t$ .

Here, $\Delta t = L/c$ corresponds to the time taken for the photon's transfer over a distance $L$ from the source to D in the lab frame.[16] It should be noted, however, that this 'reaching into the past' is not equivalent to 'the future reaching into the present', since according to PTI, there are no actualized events in the future; that is the nature of a "becoming" universe. The future consists only of quantum-level possibilities.

## References

1. Danan, A., Farfurnik, D., Bar-Ad, S. and Vaidman, L. (2013). "Asking Photons Where They Have Been," *Phys Rev Lett 111*, 240402. https://arxiv.org/abs/1304.7469.
2. Li, Z. *et al* (2013). "Comment on "Past of a quantum particle," Phys. Rev. A **88**, 046102.
3. Svennson, B. (2014). "Comments to Asking Photons Where They Have Been," https://arxiv.org/abs/1402.4315
4. Salih, H. (2015). "Comment on 'Asking Photons Where They Have Been'," *Front. Phys. 3*:47.

[14] This omission of a demonstrably crucial component, for which justification is attempted by reference to its small size, might be an appropriate topic for study in the methodology of science. When does "smallness" of a quantity render it properly negligible, and when does it not? The same issue comes up in discussions of the significance of weak values, where a nonvanishing disturbance to a measured system is often asserted to be negligible, but without that disturbance there is no information at all, so that it is nontrivial despite its smallness [35] (Kastner, 2017).

[15] One is free not to adopt that suggested ontology and can still use the transactional formulation, but it is arguably more consistent in view of the fact that quantum states are Hilbert space objects that don't "fit into" 3+1 spacetime.

[16] In the many cases in which the photon is not localized to any particular path from source to detector (i.e., not a 'which way' measurement), the time $\Delta t$ has an inherent uncertainty proportional to the lack of localization of the photon over those possible paths (i.e., uncertainty in $L$).

5. Danan, A., Farfurnik, D., Bar-Ad, S. and Vaidman, L. (2015). “Reply to a Commentary ‘Asking photons where they have been without telling them what to say’,” *Front. Phys. 30*. https://doi.org/10.3389/fphy.2015.00048; https://arxiv.org/abs/1401.5420
6. Griffiths, R. B. (2016). "Particle path through a nested Mach-Zehnder interferometer," *Phys. Rev. A 94*, 032115.
7. Sokolovski (2016). “Asking photons where they have been in plain language,” *Phys. Lett. A 381*, 227. (https://arxiv.org/abs/1607.03732)
8. Nikolaev, G. N. (2017). “Paradox of Photons Disconnected Trajectories Being Located by Means of ‘Weak Measurements’ in the Nested Mach-Zehnder Interferometer,” *JETP Lett., 105,* No. 3, pp. 152-157
9. Englert, B.-G. et al. (2017). "Past of a quantum particle revisited,"*Phys. Rev. A 96*, 022126
10. Vaidman, L. (2013). “The past of a quantum particle,” *Phys. Rev. A 87*, 052104.
11. Vaidman, L. (2017a) “A Comment on ‘Paradox of photons disconnected trajectories being located by means of 'weak measurements' in the nested Mach-Zehnder interferometer," *JETP Letters 105*, 152.
12. Vaidman, L. (2017b). “A Comment on ‘Asking photons where they have been in plain language’”, https://arxiv.org/abs/1703.03615
13. Weisniak, M. (2018). "Spectra in nested Mach–Zehnder interferometer experiments,"*Physics Letters A 382*: 36, 2565-2568
14. Y. Aharonov and L. Vaidman (2007). "The Two-State Vector Formalism of Quantum Mechanics: an Updated Review," https://arxiv.org/abs/quant-ph/0105101v2.
15. Vaidman, L. (2007) "The Two-State Vector Formalism," preprint, https://arxiv.org/abs/0706.1347.
16. Stuckey, M. (2015a) “Understanding Retrocausality,” https://www.physicsforums.com/insights/retrocausality/)
17. Stuckey, M. (2015b). “Weak Values: Part 1: Asking Photons Where They Have Been,” https://www.physicsforums.com/insights/weak-values-part-1-asking-photons/
18. Saldanha, P. (2014). “Interpreting a nested Mach-Zehnder interferometer with classical optics,” *Phys. Rev. A 89*, 033825.
19. Y. Aharonov, Bergmann, P. Lebowitz, J. (1964). “Time Symmetry in the Quantum Process of Measurement,” *Phys. Rev. 134*, B1410.
20. Bub, J. and Brown, H. (1986). “Curious Properties of Quantum Ensembles Which Have Been Both Preselected and Post-Selected,” *Phys Rev Lett 56*, 2337-2340.
21. Mermin, N. D. (1997), ‘How to Ascertain the Values of Every Member of a Set of Ob- servables That Cannot All Have Values,’ in R. S. Cohen et al. (eds), Potentiality, Entan- glement and Passion-at-a-Distance, 149-157, Kluwer Academic Publishers
22. Kastner, R. E. (1998a). “The Three-Box Paradox and other Reasons to Reject the Counterfactual Usage of the ABL Rule,” *Found. Phys. 29*, 51–863. https://arxiv.org/abs/quant-ph/9807037
23. Kastner, R. E. (2003). “The Nature of the Controversy over Time-Symmetric Quantum Counterfactuals,” *Philosophy of Science 70*, 145-163.
24. Kastner, R. E. (2004). “Weak Values and Constistent Histories in Quantum Theory,” *Stud. Hist. Philos. Mod. Phys. 35*, 57-71. https://arxiv.org/abs/quant-ph/0207182
25. Kastner, R. E. (2010). “Shutters, Boxes, but no Paradoxes: Time Symmetry Puzzles in Quantum Theory,” *International Studies in the Philosophy of Science, 18*:1, 89-94, DOI: 10.1080/02698590412331289279 Preprint: https://arxiv.org/abs/quant-ph/0207070
26. Bialynicki-Birula, I., Bialynicka-Birula, Z. (2009). "Why photons cannot be sharply localized," Physical Review A 79 (2009) 032112. https://arxiv.org/abs/0903.3712.
27. Wheeler, J. A. (2004). From the Big Bang to the Big Crunch,: (Interview), Cosmic Search Vol. 1 No. 4. http://www.bigear.org/vol1no4/wheeler.htm.
28. Cramer, J. G. (1986). The Transactional Intepretation of Quantum Mechanics. *Rev. Mod. Phys.*
29. Kastner, R. E. (2022). *The Transactional Intepretation of Quantum Mechanics: A Relativistic Treatment.* Cambridge: Cambridge University Press.
30. Kastner, R. E. and Cramer, J, G, (2018). “Quantifying Absorption in the Transactional Interpretation,” https://arxiv.org/pdf/1711.04501.pdf.
31. Price, H. (2008). “Toy models for retrocausality,” *Stud. Hist. Philos. Mod. Phys. 39*, 752-761.
32. Schlatter, A. and Kastner, R. E. (2023). Gravity from Transactions: Fulfilling the Entropic Gravity Program. *J. Phys. Commun.* **7** 065009.
33. Kastner, R. E., Kauffman, S. and Epperson, M. (2018). “Taking Heisenberg’s Potentia Seriously,” *Int'l Jour. Quan. 4*: 2, 158-172.
34. Kastner, R. E. (2018). “On the Status of the Measurement Problem: Recalling the Relativistic Transactional Interpretation,” *Int'l Jour. Quan. Foundations 4*, Issue 1, pages 128-141.
35. Kastner, R.E. (2017). Demystifying Weak Measurements. *Found Phys* **47**, 697–707. https://doi.org/10.1007/s10701-017-0085-4